\documentclass[preprint,showpacs,preprintnumbers,amsmath,amssymb]{revtex4}
\usepackage{epsfig}

\newcommand{\CaFeAs}{CaFe$_2$As$_2$}
\newcommand{\BaFeAs}{BaFe$_2$As$_2$}

\begin{document}

\title{Synthesis and Properties of CaFe$_2$As$_2$ Single Crystals}

\author{F. Ronning$^1$, T. Klimczuk$^{1,2}$, E.D. Bauer$^1$, H. Volz$^1$, J.D. Thompson$^1$}
\affiliation{$^1$Los Alamos National Laboratory, Los Alamos, New Mexico 87545, USA\\
             $^2$Faculty of Applied Physics and Mathematics, Gdansk University of Technology, Narutowicza 11/12, 80-952 Gdansk, Poland}

\date{\today}

\begin{abstract}
We report the synthesis and basic physical properties of single
crystals of CaFe$_2$As$_2$, an isostructural compound to
BaFe$_2$As$_2$ which has been recently doped to produce
superconductivity. CaFe$_2$As$_2$ crystalizes in the ThCr$_2$Si$_2$
structure with lattice parameters $a$ = 3.887(4) \AA{} and $c$ =
11.758(23) \AA{}. Magnetic susceptibility, resistivity, and heat
capacity all show a first order phase transition at $T_0$ = 171 K.
The magnetic susceptibility is nearly isotropic from 2 K to 350 K.
The heat capacity data gives a Sommerfeld coefficient of 8.2 $\pm$
0.3 mJ/molK$^2$, and does not reveal any evidence for the presence
of high frequency ($>$ 300 K) optical phonon modes. The Hall
coefficient is negative below the transition indicating dominant
n-type carriers.
\end{abstract}


\maketitle

The discovery of superconductivity in the oxypnictide compounds
RFeAs(O$_{1-x}$F$_x$) (R = La, Ce, Pr, Nd, Sm, Gd) with the ZrCuSiAs
structure type has stimulated a wealth of activity around the globe
\cite{KamiharaJACS2008, XHChenNature2008, GFChen2008a, ZARen2008a,
ZARen2008b, ZARen2008c, PCheng2008}. The similarity of this system
and the cuprates with respect to the layered structure and the phase
diagram with carrier doping, suggests that the physics may be
similar. The fact that the parent compounds are metallic indicates
additional similarity to the heavy fermion superconductors
\cite{MathurNature1998,ParkNature2006,YuanScience2003}.

If the FeAs layers are critical for the relatively high
superconducting transition temperatures then it does not come as a
big surprise that superconductivity was found in the related
ThCr$_2$Si$_2$ structure. Hole-doping by potassium in BaFe$_2$As$_2$
and SrFe$_2$As$_2$ produces superconductivity up to $T_c$ = 38
K.\cite{Rotter2008b,GFChen2008b,Sasmal2008}. Given the relatively
high transition temperatures and that single crystals appear
relatively easier to synthesize than those in the RFeAsO family, the
ThCr$_2$Si$_2$ structure may be a more ideal system for elucidating
the physics of these new Fe-based superconductors. Currently, the
AFe$_2$As$_2$ compounds are known to be stable with divalent A = Ba,
Sr, and Eu\cite{Pfisterer1980, Marchand1978}, of which only the A =
Ba and Sr compounds so far have been doped to produce
superconductivity \cite{Rotter2008b,GFChen2008b,Sasmal2008}. In this
paper we report on the synthesis and basic physical properties of
single crystals of CaFe$_2$As$_2$. A sharp first order anomaly is
observed by susceptibility, heat capacity and electrical transport
measurements at $T_0$ = 171 K.

\begin{figure}[htbp]
     \centering
     \includegraphics[width=0.5\textwidth]{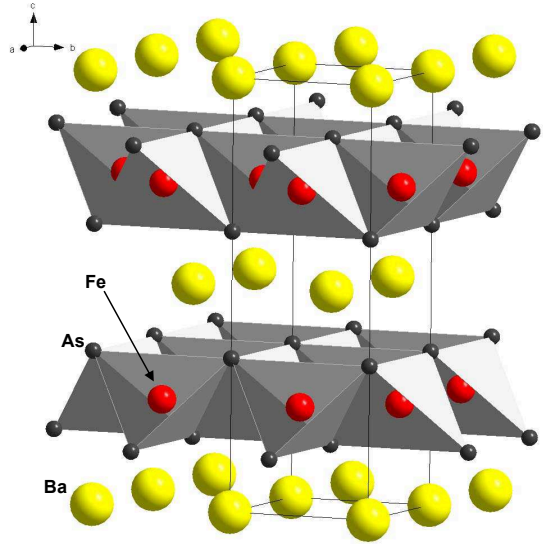}
     \caption{Crystal structure of CaFe$_2$As$_2$} \label{struc}
 \end{figure}
\begin{figure}
\includegraphics[width=3.3in]{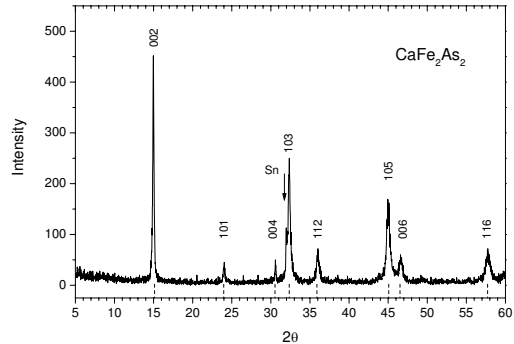}
\caption{\label{XRD} Powder X-ray diffraction pattern
(CuK$_{\alpha}$ radiation) for CaFe$_2$As$_2$. Vertical bars at the
bottom represent the Bragg peak positions for the ThCr$_2$Si$_2$
tetragonal (I4/mmm) structure with refined cell parameters a =
3.887(4) {\AA} and c = 11.758(23) {\AA}. Miller indices for each
peak are shown, and a peak from the Sn flux is marked with an
arrow.}
\end{figure}

Single crystals of CaFe$_2$As$_2$ were grown in Sn flux in the ratio
Ca:Fe:As:Sn=1:2:2:20. The starting elements were placed in an
alumina crucible and sealed under vacuum in a quartz ampoule.  The
ampoule was placed in a furnace and heated to 500 $^{\circ}$C at 100
$^{\circ}$C hr$^{-1}$, and held at that temperature for 6 hours.
This sequence was repeated at 750 $^{\circ}$C, 950 $^{\circ}$C and
at a maximum temperature of 1100 $^{\circ}$C, with hold times of 8
hr., 12 hr., and 4 hr, respectively.  The sample was then cooled
slowly ($\sim4 ^{\circ}$C hr$^{-1}$) to 600 $^{\circ}$C, at which
point the excess Sn flux was removed with the aid of a centrifuge.
The resulting plate-like crystals of typical dimensions 5 x 5 x 0.1
mm$^3$ are micaceous and ductile and are oriented with the $c$-axis
normal to the plate.  \CaFeAs{} crystallizes in the ThCr$_2$Si$_2$
tetragonal structure (space group no. 139) (Fig. \ref{struc})  with
lattice parameters $a$ = 3.887(4) \AA{} and $c$ = 11.758(23) \AA{}
as revealed by the powder x-ray diffraction pattern shown in Fig.
\ref{XRD}.  In this structure, layers of Ca are capped by Fe-As
tetrahedra along the c-axis.  These Fe-As tetrahedra are the common
structural units to the Fe-based RFeAsO and AFe$_2$As$_2$
superconductors.

Magnetic measurements were performed from 1.8 K to 300 K using a
commercial SQUID magnetometer. Specific heat measurements were
carried out using an adiabatic method in a commercial cryostat from
2 K to 300 K. Electrical transport measurements were performed using
a LR-700 resistance bridge with an excitation current of 1 mA, on
samples for which platinum leads were spot welded.

\begin{figure}[htbp]
     \centering
     \includegraphics[width=0.5\textwidth]{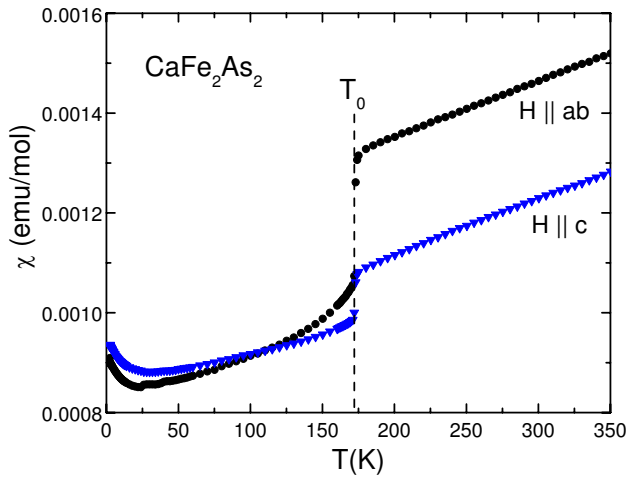}
     \caption{Magnetic susceptibility $\chi(T)$ of \CaFeAs{} measured
in a magnetic field $H=5$ T for $H||$ab (black circles) and $H||$c
(blue triangles).  A structural transition is indicated by the sharp
drop at $T_0 =172$ K (dashed line) in   $\chi_c$ and $\chi_{ab}$. }
\label{chi}
 \end{figure}

The magnetic susceptibility $\chi(T)$ of CaFe$_2$As$_2$ measured in
a magnetic field $H = 5$ T with $H||$ab and $H||$c is shown in Fig.
\ref{chi}.  The susceptibility is essentially isotropic over the
entire measured temperature range. Close to $T_0$ =172 K a sharp
drop is evident in $\chi_{ab}$ and in $\chi_c$, albeit slightly
smaller, likely indicating a structural transition that is similar
to those observed in \BaFeAs{}\cite{Rotter2008a, Ni2008,
GFChen2008b} and LaFeAsO\cite{KamiharaJACS2008}.

\begin{figure}
\includegraphics[width=3.3in]{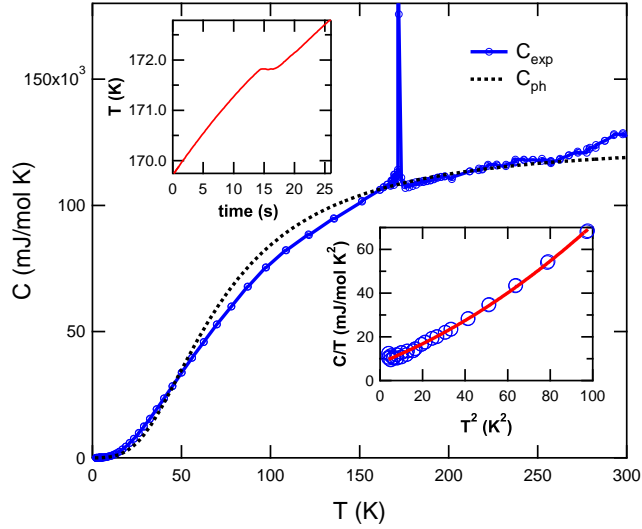}
\caption{\label{CaFeAs_Cp} Specific heat versus temperature is shown
for CaFe$_2$As$_2$. The dashed line represents a simple lattice
estimate as described in the text. The top inset displays the
relaxation curve at the transition temperature. The lower inset
displays the low temperature heat capacity. The solid line is a fit
to $C/T = \gamma + \beta T^2 + \alpha T^4$. }
\end{figure}

The heat capacity presented in figure~\ref{CaFeAs_Cp} reveals a very
sharp symmetric anomaly consistent with a first order phase
transition at 172 K (upon warming). In the top inset, the relaxation
curve of sample temperature versus time is shown. While a constant
heat is applied to the sample it steadily increases in temperature
as dictated by the sample heat capacity and the thermal link to the
bath. The plateau in the curve indicates an abrupt increase in the
heat capacity as well as the latent heat associated with the first
order transition\cite{Lashleycryogenics}, sharply defined in
temperature at 171.8 $\pm$ 0.1 K. The low temperature heat capacity
is presented in the lower inset. Below 10 K the heat capacity data
can be fit to $C = \gamma T + \beta T^3 + \alpha T^5$. This gives an
electronic specific heat coefficient of $\gamma$ = 8.2 $\pm$ 0.3
mJ/mol K$^2$. Assuming that the $T^3$ term is entirely due to
acoustic phonons, from the $\beta$ coefficient = 0.383 $\pm$ 0.018
mJ/mol K$^4$ we extract a Debye temperature $\Theta_D$ = 292 K. The
dashed curve in the figure gives the lattice contribution to the
specific heat based upon a simple Debye model using $\Theta_D$ = 292
K. While this is certainly an oversimplification of the exact phonon
density of states, the fact that this gives a reasonable account of
the data at high temperatures indicates that there are relatively
few high frequency optical phonon modes with energies above 300 K.
This is in contrast to the case of phonon-mediated superconductor
MgB$_2$\cite{Walti2001}, where analysis of the heat capacity
indicates the presence of phonon modes up to 750 K.

\begin{figure}
\includegraphics[width=3.3in]{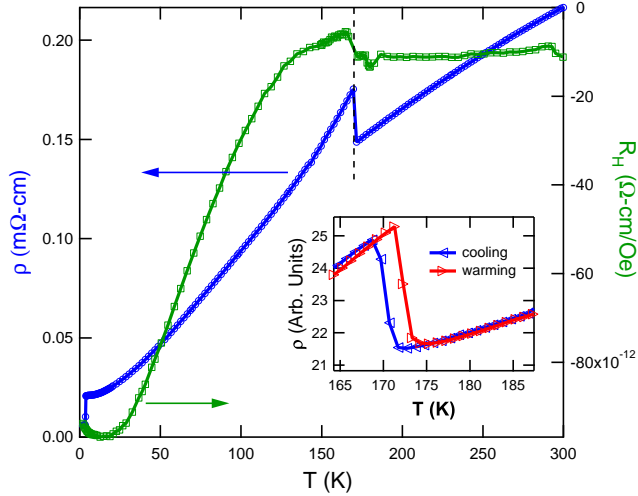}
\caption{\label{CaFeAs_Hall} In-plane resistivity (I$\parallel$ab)
and the Hall coefficient of CaFe$_2$As$_2$ as a function of
temperature. The inset illustrates the thermal hysteresis at the
transition observed in resistivity for current along the c-axis.}
\end{figure}

As with susceptibility and heat capacity, the resistivity data
presented in figure~\ref{CaFeAs_Hall} contains a clear first order
phase transition at 170 K. The jump indicates either an increase in
scattering or a decrease in the number of carriers below the
transition relative to above it. The samples have a RRR (=
$\rho$(300 K)/$\rho$(4 K)) of 10. A small partial superconducting
transition at 3.8 K is due to small Sn inclusions. Data for current
parallel to the c-axis on 4 samples (not shown) have a qualitatively
similar temperature dependence, but range in absolute magnitude from
50 to 1000 times larger than the in-plane data, possibly a
consequence of weakly coupled micaceous layers leading to large
variations in the magnitude of the c-axis resistivity. The inset
demonstrates the thermal hysteresis expected for a first order phase
transition. Also shown in Fig.~\ref{CaFeAs_Hall} is the Hall
coefficient. The dominant carrier below the 171 K transition is
electron-like. There is a role-over at 15 K which may be due to
either the multiband nature of these systems, or due to localization
effects. With a resolution of 2x10$^{-11}$ $\Omega$-cm/Oe we can not
say whether the dominant carrier type at room temperature is also
electron-like.

\begin{table}[htp]
\caption{Comparison of properties of known AFe$_2$As$_2$ compounds
where A is a divalent atom.}
\begin{ruledtabular}
\begin{tabular}{lcccccc}
Compound & $a$ (\AA) & $c$ (\AA) & $\Theta_D$ (K) & $\gamma$
(mJ/molK$^2$) & $T_{0}$ (K) & refs.
\\\hline
CaFe$_2$As$_2$ & 3.887(4) & 11.758(23)  & 292  & 8.2(3) & 171 & this work \\
EuFe$_2$As$_2$ & 3.911(1) & 12.110(4) &  &  & 195 & \cite{Marchand1978, Raffius1993, Jeevan}  \\
SrFe$_2$As$_2$ & 3.927(6) & 12.37(2) & 196 & 3.3$^1$ & 200(5) & \cite{Pfisterer1980, Yan2008, Krellner2008Sr}\\
BaFe$_2$As$_2$ & 3.962(6) & 13.04(2)  & 134,200  & 6,16,37 & 80-140 & \cite{Pfisterer1980, Rotter2008a, Ni2008, Dong2008}\\
\end{tabular}
\label{table1}
\end{ruledtabular}
$^1$ Ref \cite{Yan2008} report $\gamma$ = 33 mJ/mol K$^2$, but the
figure indicates a value between 11 and 3.3 mJ/mol K$^2$.
\end{table}


Briefly, we compare our results with currently available data on
other members of the AFe$_2$As$_2$ compounds with A = Ba, Sr, and
Eu, listed in Table 1. We note that the lattice constants
monotonically decrease from BaFe$_2$As$_2$ to SrFe$_2$As$_2$ to
EuFe$_2$As$_2$ to CaFe$_2$As$_2$, as one would expect given the
smaller ionic radii of Ca$^{2+}$ versus Eu$^{2+}$ and Sr$^{2+}$
versus Ba$^{2+}$. The higher Debye temperature for CaFe$_2$As$_2$
(292 K) is consistent with the smaller unit cell volume. However,
the structural/SDW phase transition is not monotonic with cell
volume even within the group IIA of the periodic table ranging from
80-140 K for BaFe$_2$As$_2$, and 195-205 K in SrFe$_2$As$_2$
compared with 170 K in CaFe$_2$As$_2$. (EuFe$_2$As$_2$ orders at 195
K, and the Eu moments order at 20 K\cite{Raffius1993}.) The reason
for this is not understood, but possibly indicates the sensitivity
of the transition to details of the electronic structure. Similarly,
the anisotropy of the susceptibility of CaFe$_2$As$_2$ measured at 5
T is nearly isotropic over the entire temperature range measured,
both above and below the transition. Although the overall magnitude
is similar to that observed in SrFe$_2$As$_2$\cite{Yan2008} the
anisotropy is qualitatively different. In our case the low
temperature Curie-Weiss tail is isotropic, and thus could originate
from an impurity contribution. The isotropic behavior above the
transition is more consistent with that observed in
BaFe$_2$As$_2$\cite{Ni2008}. The role of trace amounts of
ferromagnetic impurity phases such as Fe$_2$As\cite{Nowik2008}, will
be studied in more detail to determine the intrinsic behavior of the
susceptibility. Finally, whether the current samples are affected by
Sn substitution as suggested for single crystals of BaFe$_2$As$_2$
grown by a similar technique \cite{Ni2008} must still be
investigated.


We have synthesized single crystals of CaFe$_2$As$_2$, which
possesses a first order transition at 170 K, which is likely a
combined structural and magnetic transition. Given that
superconductivity has been found by doping the isostructural Ba and
Sr compounds\cite{Rotter2008b,GFChen2008b,Sasmal2008} we believe
that the Ca compound is also a likely candidate for the presence of
superconductivity upon chemical substitution.

At the completion of this work we became aware of two other papers
reporting the synthesis of CaFe$_2$As$_2$. Single crystals of
CaFe$_2$As$_2$ grown using self flux\cite{Wu2008Ca} and Sn
flux\cite{Ni2008b} methods gave results similar to ours, and indeed
superconductivity was found upon Na doping\cite{Wu2008Ca}.

\begin{acknowledgments} We acknowledge useful discussions with B. Scott.
Work at Los Alamos National Laboratory was
performed under the auspices of the U.S. Department of Energy.
\end{acknowledgments}

\end{document}